# Convolutional Recurrent Residual U-Net Embedded with Attention Mechanism and Focal Tversky Loss Function for Cancerous Nuclei Detection.

Kaushik Das (**Student ID**: 190172273)
Dept. of Electronics Engineering &
Computer Science
Queen Mary University of London
London, United kingdom (UK)
ec19142@qmul.ac.uk

Supervisor: Dr. Qianni Zhang

*Abstract* —Since the beginning of this decade, CNN (Y. LeCun et al., 1998) has been a very successful tool in the field of Computer Vision tasks. The invention of CNN was inspired from neuroscience and it shares a lot of anatomical similarities with our visual system. Inspired by the visual system anatomy of human, this paper argues that the existing naïve U-Net architecture by Olaf Ronneberger et.al. (2015) can be improved in many ways. As human visual system uses attention mechanism, we have used attention concatenation in place of normal concatenation. Although, CNN is purely feed-forward in nature but anatomical evidences show that our brain contains recurrent synapses and they often outnumber feed-forward and top-down connections. This fact inspires us to make another rectification to the U-Net model with recurrent convolution connections in place of normal convolution blocks. This paper also addresses the class imbalance issue in the field of medical image analysis that we often face. The paper resolves this problem with the help of state-of-the-art loss functions such as Dice Loss, Tversky loss and Focal Tversky loss. We have made a detailed analysis of the performances of our proposed architecture by cross-validation with the above-mentioned loss functions and demonstrated that Focal Tversky Loss Function with Recurrent Residual U-Net embedded with Attention mechanism demonstrates the most promising result. The paper demonstrates that the proposed method can be trained end to end with a few training data and it outperforms the other variant of U-Nets such as the naïve U-Net, U-Net with Recurrent Connection but without attention mechanism, U-net with attention but without recurrent connection etc. The proposed architecture has been tested primarily on Kaggle's 2018 Dataset Bowl Nuclei Segmentation Dataset and it outperforms the other variants of U-Nets already mentioned above. The statistical analysis shows advantages of the proposed architecture over other variants of U-Net models.

## I. Introduction

From the beginning of the last decade Deep Convolutional Neural Networks have surpassed many pre-existing computer vision tasks such as Object Recognition, Classification, Detection etc. Although Convolutional Neural Networks existed before this decade, but their performance was limited because of handful training data size, limited network size and limited computational capabilities. The quantum leap with AlexNet (Alex Krizhevsky at al., 2012) was primarily because of larger training dataset and using deeper network for training. Since then the success of deep convolutional neural network in the field of Computer vision is astonishing. The most common task in the field of Deep Learning is to classify an image to a particular class. However, in the field of Biomedical imaging it is necessary to label each pixel to a particular subclass e.g. for a histopathological breast cancer image it can be a requirement to find out the exact locations of the cancerous cells i.e. to label each pixel based on whether they belong to malignant or benign category. For that reason, segmentation of medical images is very important. In 2015, Olaf Ronneberger et.al. proposed a novel architecture in their U-Net paper (Olaf Ronneberger et al., 2015) to segment medical images with a very less training data. Their proposed architecture had a contracting path (Encoder) and an expanding path (Decoder) which was fully Convolutional. They proposed a copy and crop type skip connection from encoder limb to decoder limb in order to concatenate encoder layers with decoder layers that are at the same depth (Olaf Ronneberger et al., 2015). Clearly, the architecture has 3 drawbacks. Like most of the Convolutional Neural Networks, U-Net too has purely feed-forward architecture that can be thought as a rough approximation of human neuron functionality. Anatomically, it is a well-established fact that the recurrent synapses in human brain's neocortex region outnumbers feed-forward connection and these recurrent connections play a pivotal role for static object detection and segmentation tasks although in our brain object detection or recognition is a dynamic procedure (Peter Dayan et al., 2005). Hence, we have introduced recurrent convolution connections in place of normal convolution blocks. It has been experimented and concluded that human vision doesn't process an entire scene at a time instead our visual system focuses on a part of an image, video etc. i.e. it provides attention to a particular region of interest in a scene at a particular timestep and processes the data to extract meaningful information and subsequently the brain guides the eye movement to distil further information in order to make a rational decision. Moreover, the sub Networks of a Convolutional Neural Networks extracts similar low-level features repetitively. To address this issue and to resolve it, we have introduced Attention Gates (AG) (Dzmitry Bahdanau et al., 2014). Attention gates can be trained with normal recurrent CNN architectures and the AGs can learn to focus on the important feature maps without any additional guidance. At the time of testing these attention gates produces soft region proposals implicitly and focuses on the feature maps which are important for a given task. The third and most important problem regarding medical Images are high class imbalance issue i.e. there are more background information than the foreground target class e.g. malignant cells but Olaf Ronneberger et al., (2015) didn't consider this fact and their paper didn't choose the optimal loss function to address the class imbalance issue. This paper addresses the above-mentioned problem by demonstrating the repercussions of different loss functions and choose the most optimal loss function for class imbalanced dataset. We have demonstrated the necessity of the selection of proper loss function along with a stable deep-convolution-network in the light of class imbalance dataset by cross validation with Binary Cross Entropy Loss, Dice Loss (Vercauteren, T., et al., 2017), Tversky Loss (Deniz Erdogmus et al., 2017) and Focal Tversky Loss function (Nabila Abraham et al., 2019). Mathematically, as well as experimental results supports that the proposed architecture with Focal Tversky Loss function gives the best output.

## II. Background Research

Deep Convolutional Neural network has gained sky-high popularity after its state-of-the-art performance in the field of Object Recognition, Detection, Tracking, Segmentation Classification, Captioning etc. Hubel and Wiesel's experiment on cat (D. H. Hubel et al., 1959) had given a huge breakthrough to understand how we perceive objects and back then we have been observing a series of improvements in the field of computer vision such as Neocognitron (Kunihiko Fukushima 1986), HMAX (M. Riesenhuber et al., 1999), Le-Net (Yan Lecun et al., 1998) and very recently in 2012 a huge breakthrough with Alexnet (Alex Krizhevsky at al., 2012). A very deep model, VGG net was introduced (Simonyan and Zisserman 2015) which uses very small receptive field (3x3 convolution) throughout the entire architecture and the authors argued that a larger receptive field can be achieved by repetitive application of 3x3 convolution filter. Unlike AlexNet (Alex Krizhevsky at al., 2012) which uses 11x11 convolution filters at the first layer, VGG Net never uses large receptive fields. The idea of Local response normalization was presented by Alex Krizhevsky (Alex Krizhevsky at al., 2012) which normalizes response and was biologically inspired from lateral inhibition in human neuron. The idea of batch normalization was introduced (Sergey Ioffe et al., 2015) to address the problem of internal covariance shift in deep convolution Network. Batch normalization helps us to use bigger learning rate and to be reluctant about the weight initialization. It also acts as a regularization technique which can be seen as an alternative of Dropout (Nitish Srivastava et.al. 2014). It has been noticed that the usage of Batch-Normalization helps to converge the network with 14 times lesser training steps. The computational complexity of a deep convolutional neural network is very high and hence training it with CPU is an ill posed problem. Training of such deep and computationally expensive models with GPU acts as a catalyst which was first proposed by Alex Krizhevsky (Alex Krizhevsky at al., 2012) for ImageNet Large Scale Visual Recognition Challenge in 2012. Because of their innovative model architecture and training methodology the post Alexnet (Alex Krizhevsky at al., 2012) era is considered as the Renaissance era of deep learning. Ideally, a deeper model should have training error no greater than its shallower counterpart but for a very deep convolution model, it has been observed that the performance degrades and it is not because of overfitting. This problem occurs due to vanishing gradient at a later stage of a very deep model and the phenomenon is called degradation problem. The idea of residual learning has generated to nullify the degradation problem i.e. by adding previous layers activation in order to construct an identity mapping (Kaiming He et al., 2016). Semantic segmentation is a challenging field and a lot of researches have been done to map pixels of an image either as background or foreground. State-of-the-art feature extractors like Histogram of Oriented Gradients (HOG) (Navneet Dalal et al., 2005), Scale Invariant Feature Descriptors (SIFT) (David G. Lowe et al., 2004), Gist features (C. Siagian et al., 2007) works very well for computer vision tasks. Before, Deep Convolutional Neural networks era researchers used such hand-crafted features and machine learning algorithms to classify and segment images. Latest research trends have demonstrated that deep convolution neural networks produce state of the art results for image segmentation and recognition tasks (Alex Krizhevsky at al., 2012). However, deeper models used to suffer from vanishing and exploding gradients which has been prevailed by various novel ideas such as by incorporating ReLU non-linearity (Geoffrey E. Hinton et al., 2010), initialization of weights e.g. He Normal Initialization (Kaiming He et al., 2015), Xavier Initialization (Xavier Glorot et al., 2010) etc. Another idea to overcome the problem of vanishing gradient is by incorporating residual or skip connection which implements the identity mapping (Kaiming He et al., 2016). Semantic segmentation (Kevin Murphy et al., 2016) using Deep Convolution Network (DeepLab) is one of the most efficient algorithms for Image segmentation task. Another benchmark architecture for Image segmentation is SegNet (Roberto Cipolla et al., 2017) model which has an encoder-decoder like structure, where the encoder is made of 13 layers of VGG 16 network (Simonyan and Zisserman 2015) and the decoder consists of pixelwise classification layers. The decoder layers of Segnet up-samples the low dimensional feature maps. Later, an upgraded version of Segnet, known as Bayesian Segnet (Roberto Cipolla et al., 2017) was proposed that provides a Deep Learning platform for probabilistic pixel-wise semantic segmentation. However, these Deep Convolutional architectures for image segmentation are not specifically built for medical image segmentation purpose where data insufficiency and class imbalance problem are often frequent and poses a serious threat. To address these issues, Olaf Ronneberger (Olaf Ronneberger et al., 2015) proposed a novel architecture for semantic segmentation for medical images. The network consists of convolutional encoder and decoder blocks. The architecture is advantageous because of few reasons such as it allows to use global location and context at a time. The naïve U-Net model performs very well with a small training dataset. Although U-Net works very well for medical Image segmentation task, but it also works exceptionally well for general purpose image segmentation also. Several modifications of the naïve U-net have been proposed such as CNN based segmentation where 2 modifications were made. The two modifications to the naïve U-Net architecture were to combine multiple segmentation maps created at different scales and elementwise sum of forward feature maps from one part of the modified network to another (Grady Jensen et al., 2017). The momentousness of skip connection for segmentation of Biomedical Images has been studied in U-Net (Olaf Ronneberger et al., 2015) and Residual Network architecture (Michal Drozdzal et al., 2016). DCAN, another novel medical image segmentation model was proposed (Hao et.al. 2016), that uses multi-level contextual features from the hierarchical model architecture with auxiliary supervision. Fausto Milletari et.al. proposed an innovative idea to segment a 3-D medical image, known as V-net (Fausto Milletari et al., 2016) which utilizes volumetric images for segmentation. A similar idea was presented by

(Olaf Ronneberger et.al., 2016) by utilizing a variant of naïve U-net, which they termed as 3-D U-Net for 3-D medical image segmentation that learns from sparsely annotated volumetric images. Recurrent Neural Network (RNN) (Benito Fernandez et al, 1990, J. J. Hopfield, 1982, Gail a. Carpenter et al., 1987) is very famous from the beginning of its discovery for sequential data in deep learning community. Few recent state-of-the-art developments by using RNN are Image Captioning (Oriol Vinyals et al., 2015), Handwriting recognition (Alex Graves et al., 2009), Language Modelling, Neural machine Translation (Dzmitry Bahdanau et al., 2015), speech recognition (Geoffrey Hinton et al., 2013) etc.

### III. MODEL ARCHITECTURE

The architecture of the proposed model is illustrated in Figure 1 (Refer Figure 1). U Net architecture consists of a contracting path to capture context which we can call Encoder Architecture, and an expanding path or decoder to localize objects precisely (Olaf Ronneberger et al., 2015). The left limb of the proposed Recurrent Residual U-Net architecture has recurrent residual blocks as shown in figure 2.B. followed by a 2x2 max-pooling (M. Ranzato et al., 2007) layer with stride 2 for down sampling. After every max-pooling operation we double the number of feature channels as suggested (Olaf Ronneberger et al., 2015). On the other hand, the right limb of the proposed modified U-Net consists of expansive path i.e. it up-samples the features by applying Transposed Convolution, which is followed by recurrent residual blocks as depicted in figure 1 and 2.B. After applying 2D Transposed Convolution (Vincent Dumoulin et al., 2010), we increase the dimension of the image by a factor of two and at the same time we reduce the number of feature channels by a factor 2. The proposed model concatenate the feature maps from the encoding path with the up-sampled features at the right limb of the decoding path that are at the same depth by using attention mechanism (Refer Figure 1). 1x1 convolution followed by sigmoid activation is applied at the final layer to map the feature vectors to the desired output classes. The detail of each component of the proposed architecture is explained below.

**A. Attention Mechanism for U-Net:**

Here, for our architecture we have used additive Attention Gate (AG) model (Dzmitry Bahdanau et al., 2014) to focus on Cancerous Nuclei targets that poses irregular random shapes and sizes. Models trained with attention gates learns to censor regions which don't carry meaningful information while focuses on salient features for a specific task (Ozan Oktay et al. 2018). According to Ashish Vaswani et al. an attention mechanism can be defined as mapping a query and a set of key value pairs to an output. The output is computed as weighted sum of the values, where the weight assigned to each value is computed by a compatibility function of the query with the corresponding key (Ashish Vaswani et al., 2017). U nets are used for image segmentation purpose because of its proven utility and efficient use of GPU memory. At multiple scale we extract features maps which we merge later by using skip connections in order to combine coarse and fine level dense prediction (Ozan Oktay et al. 2018) (Figure 1). Attention coefficient $\alpha_i \in [0,1]$ identify the salient features and puts more weight on them and prunes un-necessary feature responses to keep only the relevant activations for a specific task. The output of an attention gate is the elementwise product between the attention coefficient and the features and can be formularized as $\hat{x}_y^l = x_y^l \cdot \alpha_i^l$. The term $\alpha_i^l$ is the attention coefficient of $x_i^l$ where $x_i^l \in \mathbb{R}^{F_l}$ and $F_l$ represents the number of feature maps in layer l (Ozan Oktay et al. 2018). To determine the area of interest, we use gating vector $g_i \in \mathbb{R}^{F_g}$ on each pixel. This gating vector prunes lower level feature responses. Here for our project we have used additive attention to construct the gating vector. Additive attention can be formulated as:

$$q_{att}^x = \psi^T \left( \Phi \left( U_{att}^T S_t^l + W_{att}^T h + b_g \right) \right) + b_\psi \quad (1)$$

$$\alpha_i^l = \sigma(q_{att}^x(S_t^l, h; \varphi)) \quad (2)$$

Where $\sigma(x; \varphi) = \frac{1}{1+e^{-\varphi^T x}}$ defines the sigmoid activation function. Attention gate is made of a set of parameters $\varphi$ which is a combination of linear transformation of $U_{att}^T \in \mathbb{R}^{DXK}$ and $W_{att}^T \in \mathbb{R}^{DXK}$ and $\psi^T \in \mathbb{R}^D$ and the bias term $b_\psi \in \mathbb{R}$ and $b_g \in \mathbb{R}^{DX1}$. In the above equation the decoder state at layer l is denoted by $S_t^l$ and the encoder state is denoted by h. We have taken $\Phi$ as a non-linear activation function, in our case it is tan hyperbolic function. Between Rectified Linear Unit and Tan Hyperbolic non-linear activation function, the latter shows more promising results. For neural machine translation or image captioning task we use $\sigma$ as SoftMax activation function in order to normalize the attention coefficient in equation 2. However, sequential applications of SoftMax activation function demonstrates sparsity at the output. Hence, we have used sigmoid activation function in equation 2 (Ozan Oktay et al. 2018). These attention gates can be integrated with the proposed U-Net architecture by Olaf Ronneberger et.al. in place of direct concatenation layer (in the original paper which was referred as copy and crop connection) through skip connections in order to focus only on the relevant salient features (Olaf Ronneberger et al., 2015). Information extracted at the different scale are multiplied with attention coefficients which is specially very important at the coarser scale as these layers produce naïve, noisy and irrelevant features which are nullified by attention mechanism. Unlike naïve U-Net's (Olaf Ronneberger et al., 2015) concatenation our model filters, keeps and concatenates only the relevant activations through skip connections. These Attention Gates refine the activations both during forward and backward propagation. During back-propagation the attention gates learns the relevant features and down-weights the irrelevant features. Hence, the model's layers update the parameters based on the spatial regions which are pertinent for our nuclei segmentation task. For j[th] layer, mathematically the update rule can be formulated as below:

$$\frac{\partial(\hat{x}_i^{j+1})}{\partial(\varepsilon^j)} = \frac{\partial\left(\alpha_i^{j+1} \cdot f\left(x_i^j; \varepsilon^j\right)\right)}{\partial(\varepsilon^j)} = \frac{\alpha_i^{j+1} \partial f\left(x_i^j; \varepsilon^j\right)}{\partial(\varepsilon^j)} + \frac{\partial\left(\alpha_i^{j+1}\right)}{\partial(\varepsilon^j)} x_i^{j+1} \quad (3)$$

Image patches in j[th] layer must produce relatively meaningful or mature features with respect to our nuclei segmentation task. Hence, it is expected to obtain the effectiveness of the Attention mechanism at the later stage of the pipeline. Moreover, different features are available at different scales. Hence, in order to learn the diverse and complementary attention coefficient Saumya Jetley et.al. proposed to employ attention over different spatial resolutions (Saumya Jetley et al., 2018). The above intuitions led us to use attention mechanism after the Convolution blocks at the later stage in the pipeline which needs to be done before Max-pool Operation to ensure that there is no fall in spatial resolution (Saumya Jetley et al., 2018). Hence, we haven't used Attention Gate for the first depth concatenation, instead we have used Normal skip connection (referred as copy and crop in Olaf Ronneberger et al., 2015) to concatenate because the extracted features at such a shallow level is not 'matured' and they don't represent features in high dimensional space.

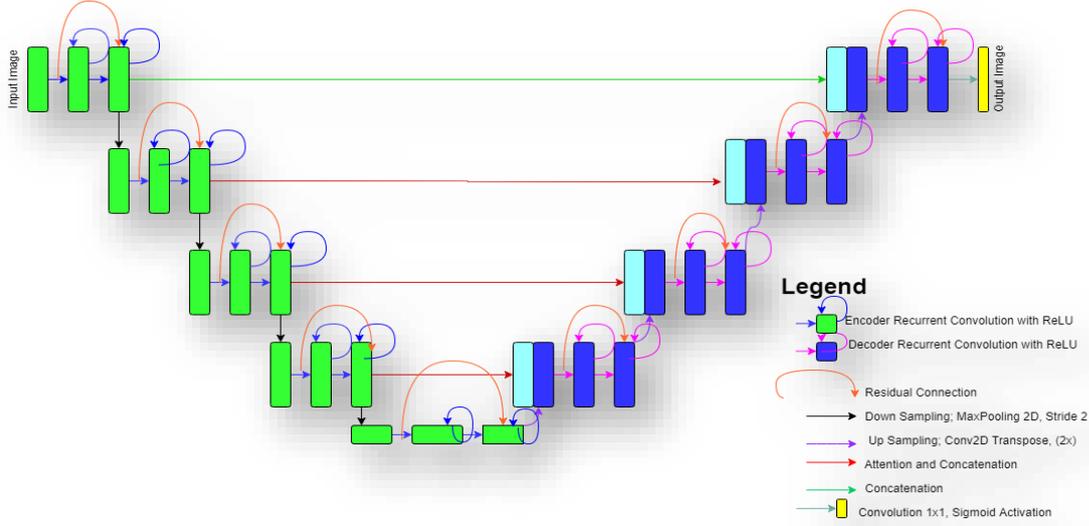

Figure 1: Architecture of the proposed Recurrent Residual Attention U-Net Model.

**B. Recurrent Residual Convolution Connection:**

The main ingredient of our proposed architecture is recurrent connection. The recurrent architecture has been implemented in a way so that the blocks evolve over the discrete timestep as proposed in (Ming Liang et al., 2015). Let us assume the empirical notation of an input unit of the Recurrent Block at $l^{th}$ layer is $x^l$ and there exist a pixel at $j^{th}$ feature map whose pixel co-ordinate is (m,n). In addition, let's consider the output of the layer is $Z^l_{m,n,t}(t)$ at timestep t. Mathematically, $Z^l_{m,n,t}(t)$ can be formulated as:

$$Z^l_{m,n,t}(t) = \left(\theta^g_j\right)^T * x^{g(m,n)}_l(t) + \left(\theta^r_j\right)^T * x^{r(m,n)}_l(t-1) + b_l \quad (4)$$

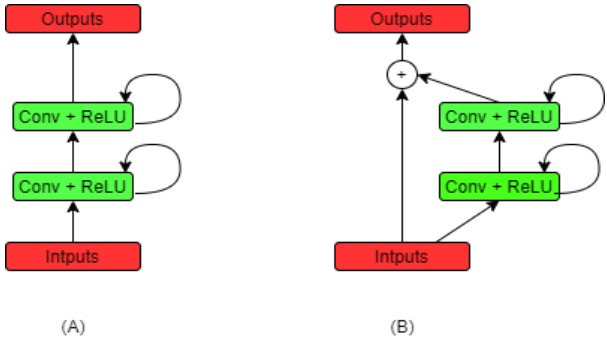

Figure 2. (A) Recurrent Convolution Unit. (B) Recurrent Residual Convolution Unit.

Where $x^{g(m,n)}_l(t)$ denotes the normal feedforward input whereas $x^{r(m,n)}_l(t-1)$ represents the recurrent input that are vectorized patches and centred at (m,n) of the feature maps at current and previous layers respectively (Chris Yakopcic et al., 2018). $\theta^g_j$, $\theta^r_j$ are the feedforward and recurrent weights respectively and $b_l$ denotes the bias term. The output of the Recurrent convolutional layer is then processed as:

$$p^l_{m,n,t}(t) = h(f(Z^l_{m,n,t})) \quad (5)$$

Here, f denotes non-linear activation function and this paper uses ReLU non-linearity and it can be mathematically formulated as $f(Z^l_{m,n,t}) = \max(Z^l_{m,n,t}, 0)$ (Geoffrey E. Hinton et al., 2010) Whereas h represents Batch-Normalization (BN) (Sergey Ioffe et al., 2015) function in order to prevent the network suffering from internal-co-variate shift and it can be formulated as below: For the brevity of notation we omit timestep (t) while calculating mean and variance in the below notations:

$$\mu^l = \frac{1}{m}\sum f(Z^l_t) \quad \text{Mini-batch mean (6.a)}$$

$$\sigma^2_l = \frac{1}{m}\sum (f(Z^l_t) - \mu^l)^2 \quad \text{Mini-batch variance (6.b)}$$

$$\hat{x}^l_t = \frac{f(Z^l_t) - \mu^l}{\sqrt{\sigma^2_l + \epsilon}} \quad \text{Normalization (6.c)}$$

$$h = \lambda \hat{x}^l_t + \beta \equiv BN_{\lambda,\beta}\left(f(Z^l_t)\right) \quad \text{Scale and Shift (6.d)}$$

Internal Co-Variate shift is a phenomenon in deep neural network where the change in distribution of network activation due to frequent change of network parameters during training time. Batch normalization prevents the internal co-variate shift of neural network (Sergey Ioffe et al., 2015). BN allows us to use bigger learning rates and to be less careful about weight initialization. It acts as a regularization function which prevents overfitting and the necessity of dropout can be shut out (Sergey Ioffe et al., 2015). Following the principal proposed by Wiesler and Ney which claims DNN converges faster if the inputs are whitened, we use batch normalization for better and faster convergence (Simon Wiesler et al., 2011). Equation 6.c whitens the inputs of each stand-alone network and $\epsilon$ prevents exploding the inputs when the mini-batch variance is zero in equation 6.b. Equation 6.d provides identity transformation to the whitened input distribution (zero mean and unit variance) so that we don't constrain the stand-alone networks' inputs to work on a fixed regime of non-linear activation functions.

For residual architecture we add up the Recurrent Convolution Output and Residual unit as shown in figure 2.B. Mathematically, if the output of the recurrent convolution block is expressed as $Y_1$ then the output of the recurrent residual connection can be formulated as:

$$Z_t^{l+1} = Z_t^l + h(f(Z_t^l)) \quad (7)$$

The outputs of these Recurrent Residual Convolution layers are used as inputs to the next layer as depicted in Fig 1. This residual architecture addresses the degradation problem of deeper network by introducing deep residual learning framework. This expression in equation 7 can be realized by feedforward neural network with shortcut connection / Skip Connection (Kaiming He et al., 2016). The idea of residual learning has generated to nullify the degradation problem i.e. the added layers can be constructed as identity mappings, which ensures a deeper model should have training error no greater than its shallower counterpart (Kaiming He et al., 2016).

## IV. LOSS FUNCTION

Fully convolutional neural network achieves excellent accuracy for Image segmentation tasks (Olaf Ronneberger et al., 2015). But, the main challenge towards segmentation of medical image is due to high data imbalance e.g. for lesion segmentation or Nuclei Segmentation the number of nuclei/lesion voxels are often much lesser than the non-nuclei/non-lesion voxels. Training with such kind of unbalanced dataset is quite problematic in the field of medical domain because this kind of data will lead us towards high precision and low recall/ Sensitivity. In medical domain false negative is much less desirable than false positive (Deniz Erdogmus et al., 2017). Weighted binary cross-entropy loss tried to solve this problem by assigning weights to the imbalanced class and it can be written as

$$\text{WBCE} = -[\beta y \log(\hat{y}) + (1-\beta)(1-y)\log(1-\hat{y})] \quad (8)$$

Where y signifies original class and $\hat{y}$ is the predicted label of the pixels of an Image. Hence, by controlling the value of weight ($\beta$) we can solve the problem of imbalance data. However, it is very difficult to figure out the proper value of $\beta$ for segmentation task. Hence, we looked into another loss function which is commonly known as Dice Loss (Tom Vercauteren et al., 2017). Dice Loss of two class variant can be defined as –

$$DL_2 = \frac{\sum_{n=1}^N (p_n g_n) + \epsilon}{\sum_n P_n + g_n + \epsilon} + \frac{\sum_{n=1}^N (1-p_n)(1-g_n) + \epsilon}{\sum_n 2 - P_n - g_n + \epsilon} \quad (9)$$

Let R be the reference foreground segmentation (gold standard) with voxel values $g_n$, and P the predicted probabilistic map for the foreground label over N image elements $p_n$, with the background class probability being 1 – $P_n$ (Tom Vercauteren et al., 2017). $\epsilon$ in equation 9 prevents instability of the loss function i.e. it prevents dividing by 0 when g and p are empty. The dice similarity coefficient (D) can be written more formally as-

$$D = \frac{2|pG|}{|p| + |G|} = \frac{2TP}{2TP + FP + FN} \quad (10)$$

Where P and G are the predicted and ground truth binary labels of an image. From the above equation we can see that Dice coefficient weights false positive and false negative equally which results in segmentation maps with high precision score but low recall value. But, in the field of medical image analysis it is very important to reduce false negative rate. The paper argues to put more weight on False Negative (FN) than false Positive (FP) during training so that the network can learn to detect small nuclei. Hence, we introduce another loss function based on Tversky Index (Deniz Erdogmus et al., 2017). Tversky Index T (α, β) is conceptually a slight modification of Dice Coefficient and it can be formulized as below –

$$T(\alpha, \beta) = \frac{\sum_{i=1}^N p_{0i} g_{0i}}{\sum_i p_{0i} g_{0i} + \alpha \sum_i p_{0i} g_{1i} + \beta \sum_i p_{1i} g_{0i}} \quad (11)$$

In the above equation $\alpha$ and $\beta$ controls the magnitudes of penalty of false positive and false negative respectively. The Sigmoid output layer of the network either gives $p_{0i}$ which is the probability that voxel i belongs to a nucleus or $p_{1i}$ that indicates the probability that voxel i doesn't belong to a nucleus. Also $g_{0i}$ is 1 for a nuclei voxel and 0 for a non-nuclei voxel, similarly $g_{1i}$ is 1 for non-nuclei voxel and 0 for a nuclei voxel. The Tversky Index T $(\alpha, \beta)$ adapted to a loss function by minimizing $\sum(1 - T(\alpha, \beta))$.

The gradient of the loss can be written as: $\frac{\partial T(\alpha,\beta)}{\partial P_{0i}} =$

$$\frac{g_{0j}\left(\sum_i p_{0i}g_{0i} + \alpha \sum_i p_{0i}g_{1i} + \beta \sum_i p_{1i}g_{0i}\right) - (g_{0j} + \alpha g_{1j})\sum_{i=1}^N p_{0i}g_{0i}}{\left(\sum_i p_{0i}g_{0i} + \alpha \sum_i p_{0i}g_{1i} + \beta \sum_i p_{1i}g_{0i}\right)^2} \quad (12)$$

$$\frac{\partial T(\alpha,\beta)}{\partial P_{1i}} = \frac{-\beta(g_{0j})\sum_{i=1}^N p_{0i}g_{0i}}{\left(\sum_i p_{0i}g_{0i} + \alpha \sum_i p_{0i}g_{1i} + \beta \sum_i p_{1i}g_{0i}\right)^2} \quad (13)$$

The trade-off between the weights of false negative and False Positive can be controlled by adjusting the value of hyperparameters $\beta$ and $\alpha$ respectively. If we set the value of α = 0.5 and β = 0.5 the Tversky Coefficient in equation 11 will be equivalent to Dice Coefficient in equation 10 which is analogous to F1 score. If we vary the value of α and β in such a way so that (α + β) =1, then we will get the set of $F_\beta$ scores. Setting higher value of β ensures enforcing more weight on False Negative (FN)/recall value than Precision. Mathematically from equation 11 we can claim that higher value of β for training a Network may lead to better generalization which can eventually bypass the problem of imbalanced dataset that in due course boosts recall (Deniz Erdogmus et al., 2017).

We have partially solved the issue with imbalanced data for which the network struggles to segment small ROIs successfully as they don't contribute significant loss. Hence, in order to make the loss significant even for small highly imbalanced dataset, we have introduced Tversky Index that puts different weights for the False Positive (FP) and False Negative (FN) ratios. We can still improve the Tversky loss function by accepting the fact that, the Image Background is easy example whereas small ROIs are difficult training examples. This observation introduces the concept of focal loss that down-weights the loss for well classified / easy examples whereas it imposes higher penalty for hard misclassified small ROIs (Priya Goyal et al., 2017). The focal Tversky loss (FTL) can be defined as below:

$$FTL(\alpha, \beta, \gamma) = (1 - T(\alpha, \beta))^{\gamma} \qquad (14)$$

Here $T(\alpha, \beta)$ is the Tversky coefficient and is defined in equation (11) above (Nabila Abraham et al., 2015). The FTL focusses more on misclassified ROIs i.e. For any $0 < \gamma < 1$, intuitively we can say if a pixel is misclassified with high Tversky Index (TI) then the Focal Tversky loss (FTL) is almost unaffected whereas for a misclassified pixel with lower TI, the FTL value will be heavily affected. It can be seen that the FTL $(\alpha, \beta, \gamma)$ reinforces more penalty for a misclassified hard ROI than normal Tversky loss $T(\alpha, \beta)$. For $\alpha = \beta = 0.5$ and $\gamma = 1$ the FTL in equation 14 is equivalent to Dice Loss. In the above Focal Tversky Loss equation $\alpha, \beta$ stabilize the class imbalance whereas the hyperparameter $\gamma$ determines the difficultness of a misclassified region of interest.

## V. MODEL TRAINING

### A. Image Preprocessing:

To train our model we take the dataset from Kaggle's Dataset Bowl Challenge 2018 (Booz Allen Hamilton & Kaggle, 2018) and the dataset is still available on Kaggle's website. The training data is a set of 640 Nuclei Images which were acquired under a variety of conditions and vary in the cell type, magnification, and imaging modality (brightfield vs. fluorescence). The dataset for this paper contains RGB H&E images of cancerous cell nuclei and their respective masks. For any H&E image the nuclei are labelled separately as it's masks i.e. if an H&E image contains 50 nuclei, then the dataset has 50 different masks for that image and each mask represents the location of a single nucleus. First, we have transformed all the RGB image and their respective masks to their grayscale equivalent and then we have combined all the masks of an image together so that we can visualize them as a single image. This step is important because later we will feed our Deep Neural Network an Image and its mask simultaneously to train the model. To combine individual masks of an image, we first converted all its masks to binary images and then performed Logical OR operation in order to stitch all the separated nuclei together. Once we finish this concatenation process, we convert the concatenated masks into grayscale level again. Data augmentation (A. Mikołajczyk et al., 2018) is a very useful technique here to train the model data invariance and robustness as we have very less training dataset and their respective segmented target labels. Hence, we have applied data augmentation techniques such as skewing, rotating, rescaling, zooming, flipping, shearing, height and width shifting for training the. Random elastic deformation has been considered as a very powerful way to tarin such networks when there are very limited number of training annotated examples.

### B. Training Methodology:

The proposed Recurrent Residual Attention U-Net with Focal Tversky loss has been benchmarked against the popular U-Net architecture. We have primarily used Kaggle's Nuclei segmentation dataset (Booz Allen Hamilton 2018) for training(640 images) the proposed model. We have resized the original training as well as target images to 256x256x1 dimensional images. For a very deep convolution neural network such as ours, the weight initialization is very important else a part of the network may give excessive activation and other parts may not contribute much (Olaf Ronneberger et al., 2015). Our intuition is to draw weights so that each feature map has ideally Unit Variance. For our architecture with ReLU (Geoffrey E. Hinton et al., 2010) non-linear activation function the above mention objective can be achieved by picking up weights from a Gaussian Distribution with $\frac{2}{\sqrt{N}}$ standard deviation where N denotes the total number of incoming nodes to one neuron (Kaiming He et al., 2015). We have studied 19 different combinations of the proposed architecture and different loss functions e.g. Binary Cross Entropy (BCE) Loss, Dice Loss (DL), Tversky Loss (with different hyperparameter α, β settings), Focal Tversky Loss (with different hyperparameter α, β, γ settings) and a combination of BCE and Dice Loss. Ablation cross-validation outcome are recorded in Table 1 for different statistics like Dice Coefficient, Precession, Recall, Accuracy, Area Under the Curve (AUC) and Cohen Kappa. We have taken Cohen Kappa into consideration in order to find the agreement between the Radiologist who segmented the nuclei and our proposed architecture. The one-fold cross validation statistics blended with the proposed network architecture shows that the Focal Tversky Loss function with hyperparameter combination α=0.3, β=0.7, γ = 0.75 results in the maximum Dice Coefficient, Precision, recall, Accuracy, AUC and Cohen kappa Statistics. We have found a very close result with hyper-parameter combination α=0.40, β=0.60, γ = 0.75. For almost every case the focal Tversky loss function gives an accuracy score of 0.90 around. But, for medical image analysis it is very important to get a higher recall value than precession and hence we have selected a hyperparameter combination α=0.30, β=0.70, γ = 0.75. For rest of the training process we have used this hyperparameter combination with Focal Tversky loss function. Table 1 shows the validation data-set results. As we are more interested to reduce the false negative ratio hence, we didn't tune the value of the hyperparameter α beyond 0.50 and the value of hyperparameter β below 0.50. For α = β = 0.5 and γ = 1 the FTL is equivalent to Dice Loss. The Kaggle Nuclei dataset Segmentation Dataset Experiment (Booz Allen Hamilton & Kaggle, 2018) was trained for 20 epochs with a batch-size of 4 images. We have used pixelwise Focal Tversky loss function (Nabila Abraham et al., 2015) to penalize the final output feature map as described in Equation 14. The model was optimized with Adam optimizer (Diederik p. Kingma et al., 2015) with Nesterov Accelerated Gradient (Aleksandar Botev et al., 2017).

| Loss Function | Hyperparameter α, β, γ | Dice Coefficient | Precision | Recall | Accuracy | ROC AUC | Cohen Kappa |
|---|---|---|---|---|---|---|---|
| Focal Tversky | α=0.4, β=0.6, γ = 0.75 | 0.71 | 0.74 | 0.78 | 0.91 | 0.86 | 0.67 |
| **Focal Tversky** | **α=0.3, β=0.7, γ = 0.75** | **0.75** | **0.79** | **0.81** | **0.92** | **0.88** | **0.73** |
| Focal Tversky | α=0.2, β=0.8, γ = 0.75 | 0.65 | 0.74 | 0.63 | 0.88 | 0.78 | 0.59 |
| Focal Tversky | α=0.3, β=0.7, γ = 0.80 | 0.72 | 0.79 | 0.74 | 0.91 | 0.84 | 0.67 |
| Focal Tversky | α=0.3, β=0.7, γ = 0.90 | 0.58 | 0.75 | 0.5 | 0.82 | 0.71 | 0.48 |
| Focal Tversky | α=0.3, β=0.7, γ = 0.65 | 0.67 | 0.67 | 0.76 | 0.9 | 0.84 | 0.62 |
| Focal Tversky | α=0.2, β=0.8, γ = 0.65 | 0.71 | 0.81 | 0.66 | 0.9 | 0.81 | 0.65 |
| Focal Tversky | α=0.2, β=0.8, γ = 0.55 | 0.73 | 0.84 | 0.71 | 0.91 | 0.83 | 0.68 |
| Focal Tversky | α=0.2, β=0.8, γ = 0.45 | 0.52 | 0.79 | 0.41 | 0.71 | 0.66 | 0.38 |
| Focal Tversky | α=0.2, β=0.8, γ = 0.50 | 0.45 | 0.81 | 0.33 | 0.59 | 0.59 | 0.26 |
| Focal Tversky | α=0.2, β=0.8, γ = 0.85 | 0.32 | 0.88 | 0.2 | 0.39 | 0.51 | 0.09 |
| Focal Tversky | α=0.4, β=0.6, γ = 0.85 | 0.72 | 0.73 | 0.79 | 0.91 | 0.86 | 0.69 |
| Focal Tversky | α=0.4, β=0.6, γ = 0.65 | 0.51 | 0.68 | 0.43 | 0.76 | 0.67 | 0.38 |
| Dice Loss | Not Applicable | 0.51 | 0.46 | 0.76 | 0.88 | 0.83 | 0.47 |
| BCE-Dice-Loss | Not Applicable | 0.5 | 0.7 | 0.41 | 0.79 | 0.67 | 0.4 |
| Tversky | α=0.2, β=0.8 | 0.72 | 0.84 | 0.68 | 0.9 | 0.82 | 0.65 |
| Tversky | α=0.3, β=0.7 | 0.71 | 0.75 | 0.74 | 0.9 | 0.84 | 0.67 |
| Tversky | α=0.4, β=0.6 | 0.57 | 0.54 | 0.82 | 0.89 | 0.86 | 0.53 |
| Binary Cross Entropy | Not Applicable | 0.67 | 0.74 | 0.65 | 0.88 | 0.79 | 0.6 |

Table 1: Cross validation result for different loss function with different hyperparameter combination with the proposed model.

which is known as NADAM (Timothy Dozat, 2016). The model was trained in Keras with Tensorflow backend having an initial learning rate .0001 and decaying factor $10^{-5}$ after every epoch. The training was done by using a 12GB NVIDIA Tesla K80 GPU on Google Colab Platform. We have optimized the parameters using grid search method (James S Bergstra et al., 2011). We have incorporated Checkpoint Ensembles (Hugh Chen et al., 2017) while training, so that if the validation score or accuracy of the model fails to improve after a set of epochs, the combination of weights for which the model achieved the best validation statistics, becomes the learned parameters of the model and eventually that weights will be used to test the model. Checkpoint Ensemble provides the advantages of ensembling within a single network architecture during the training process by the help of validation statistics. Checkpoint Ensemble reduces the risk of overfitting and at the end of training process the deep model doesn't end up exactly in the optimum point in the parameter space (Hugh Chen et al., 2017). With relatively higher learning rate the model explores the solution space more recklessly than smaller learning rate and hence it expects to explore more valleys (Optima) in the solution space. As higher learning rates needs lesser epochs to converge to the optima (Consider without being stuck at local minima) hence Checkpoint Ensemble can be thought as a catalyst to help the neural network to find its global optima. This paper considers another approach for better convergence during training by reducing learning rate when the optimizer was on plateau. In this situation the model performance didn't improve over few epochs that indicated the optimizer was on some plateau. During this situation, in order to improve the performance, the optimizer should move towards the error surface i.e. towards it's minima and hence we had reduced the learning rate. We realized the process was advantageous during the training. We reduced learning rate only for few epochs when it was stuck at a flat region of the error surface.

## VI. EXPERIMENTAL RESULT

The primary objective of this experiment is to demonstrate the effectiveness of the proposed model on Kaggle Nuclei Segmentation Dataset Experiment (Booz Allen Hamilton & Kaggle, 2018). An example of the dataset, it's segmented mask and our output is given in Image 3. Each image comes with fully annotated ground truth nuclei segmented labels. As the ground truth of the test dataset is not available yet so for calculation purpose, we had kept aside 30 images and their ground truth segmented labels from the training dataset. Although we have used Recurrent Residual U Net with attention Mechanism for this research work, but for a comparative study we have considered two other variants of this model e.g. Recurrent U Net with Attention Mechanism and Recurrent Residual U-Net. In order to show that our proposed model doesn't particularly perform well on a specific database, we took another benchmark dataset, which is known as 2D EM Segmentation Challenge Dataset (Albert Cardona et al., ISBI 2012 Conference). The dataset contains a full stack of EM slices to train the proposed architecture. It contains data from real world with some noise and small image alignment errors. The training data is a set of 25 sections from a serial section Transmission Electron Microscopy (ssTEM) data set of the Drosophila first instar larva ventral nerve cord (VNC). The microcube measures 2 x 2 x 1.5 microns approx., with a resolution of 4x4x50 nm/pixel (Albert Cardona et al., ISBI 2012 Conference). The test dataset consists of 5 datasets from the same Drosophila first instar larva VNC. The ISBI 2012 dataset underwent the same image preprocessing and the exact same training strategy as mentioned in section V.A and V.B respectively. An example of the dataset, it's segmented mask and our output is given in Image 4. Based on the statistics gathered, we can see from Table 2 that the proposed model with Focal Tversky loss gives the best Dice Coefficient Value (0.82 for Data Science Bowl and 0.86 for ISBI 2012 EM Challenge), Precision (0.93 for Data Science Bowl and 0.99 for ISBI 2012 EM Challenge) and Recall(0.76 for both the Dataset), Accuracy (0.94 for Data Science Bowl and 0.81for ISBI2012 EM Challenge) and ROC-AUC (0.87 for Data Science Bowl and 0.99 for ISBI 2012 EM Challenge) which is better than the other two architectures. Hence, our proposed architecture is better than any other variants of U-Net Model.

| Model | Loss Function | Hyperparameter α,β,γ | Dice Coefficient | Precision | Recall | Accuracy | ROC AUC | Dataset |
|---|---|---|---|---|---|---|---|---|
| Recurrent Residual UNET + Attention | Focal Tversky | α=0.3, β=0.7,γ = 0.75 | 0.82 | 0.93 | 0.76 | 0.94 | 0.87 | Data Science Bowl 2018 |
| Recurrent Residual UNET + Attention | Focal Tversky | α=0.3, β=0.7,γ = 0.75 | 0.86 | 0.99 | 0.76 | 0.81 | 0.88 | EM ISBI 2012 |
| Recurrent UNET + Attention | Focal Tversky | α=0.3, β=0.7,γ = 0.75 | 0.79 | 0.89 | 0.72 | 0.93 | 0.84 | Data Science Bowl 2018 |
| Recurrent UNET + Attention | Focal Tversky | α=0.3, β=0.7,γ = 0.75 | 0.83 | 0.98 | 0.71 | 0.75 | 0.85 | EM ISBI 2012 |
| Recurrent Residual UNET | Focal Tversky | α=0.3, β=0.7,γ = 0.75 | 0.77 | 0.94 | 0.67 | 0.91 | 0.83 | Data Science Bowl 2018 |
| Recurrent Residual UNET | Focal Tversky | α=0.3, β=0.7,γ = 0.75 | 0.83 | 0.99 | 0.71 | 0.74 | 0.85 | EM ISBI 2012 |

Table 2: Test output Statistics for both Data Science Bowl and EM ISBI challenge dataset for different models.

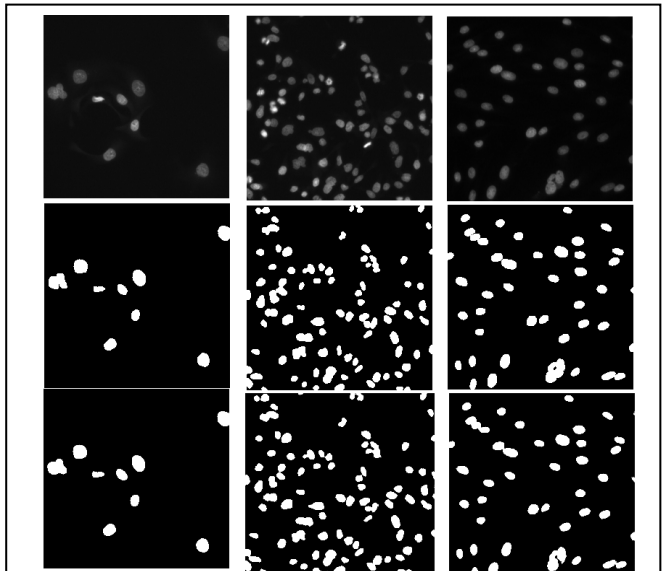

Figure 3: Experimental outputs of Data Science Bowl 2018 dataset by using the proposed architecture: First row indicates the given images; second row is the ground truths and third row contains the outputs from the proposed architecture.

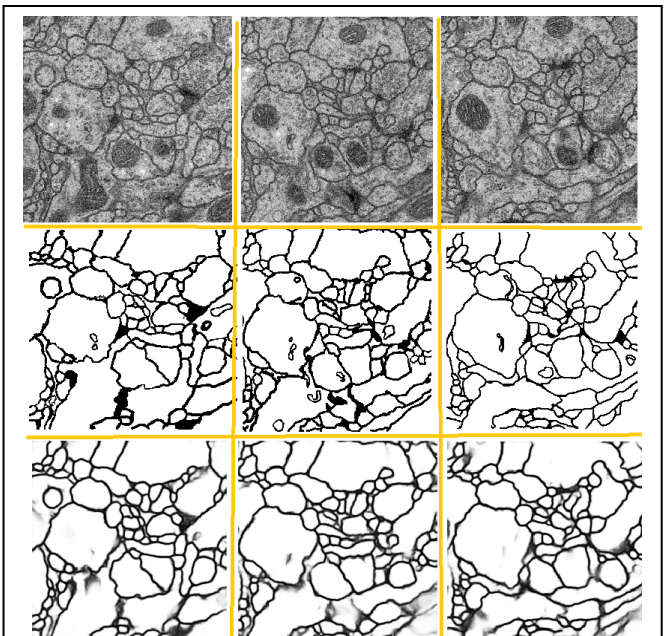

Figure 4: Experimental outputs of ISBI 2012 EM Segmentation Challenge dataset by using the proposed architecture: First row indicates the given images; second row is the ground truths and third row contains the outputs from the proposed architecture.

## VII. CONCLUSION AND FUTURE WORK

The proposed U-Net architecture with Recurrent Residual connection embedded with attention mechanism with Focal Tversky loss function demonstrates the best performance on different bio-medical datasets as shown above in Table 2. Introduction of Recurrent connection is bio-inspired because in our brain we have more recurrent synapses for static object detection task. Experimental results in Table 2 demonstrate that recurrent connection with residual blocks and attention mechanism give the best result for object detection. The implementation of the proposed network is heavily influenced by computational neurobiology and human psychology. Statistically, we can see that the performance has been boosted significantly as we consider the anatomy of human visual system and mimic the same in our architecture. The data augmentation technique played a pivotal role as we had very limited annotated datasets for training. It would be interesting to observe how many more facts of computational-neurobiology and cognitive psychology e.g. attention impact the advancement of deep-learning for computer vision task in near future. In the future, I would like to upgrade the proposed architecture with more neuro-biological facts about human vision system.

## VIII. ACKOWLEDGEMENT

I thank Professor Dr. Qianni Zhang for helpful discussions. I thank Google for providing GPUs to train the model.